\documentstyle[pra,aps,amssymb,graphicx,floats]{revtex}
\begin{document}

\draft
\wideabs{

\title{The kinetic and interaction energies of a trapped Bose gas:
Beyond the mean field}

\author{A. Yu. Cherny$^{1,2}$ and A. A. Shanenko$^{2}$}

\address{$^1$Max-Planck-Institut f\"ur Physik komplexer Systeme,
N\"othnitzer Strasse 38, D-01187 Dresden, Germany\\
$^2$Bogoliubov Laboratory of Theoretical Physics,
Joint Institute for Nuclear Research, 141980, Dubna, Moscow region,
Russia}

\date{November 30, 2001}

\maketitle

\begin{abstract}
The kinetic and interaction energies of a three-dimensional (3D)
dilute ground-state Bose gas confined in a trap are calculated beyond
a mean-field treatment. They are found to depend on the pairwise
interaction trough two characteristic lengths: the first, $a$, is the
well-known scattering length and the second, $b$, is related to the
latter by $b=a-\lambda\partial a/\partial \lambda$ with $\lambda$
being the coupling constant. Numerical estimations show that
the pairwise interaction energy of a
dilute gas of alkali atoms in a trap is negative (in spite of the
positive scattering length); its absolute value is found by about
the order of magnitude larger than that of 
the mean-field interaction energy that corresponds to
the last term in the Gross-Pitaevskii (GP) functional.
\end{abstract}
}


It is known~\cite{GP,RMP} that the total energy $E$ of a dilute 3D
Bose gas confined in a trap can be represented at $T \ll T_{c}$
($T_{c}$ is the temperature of the Bose-Einstein condensation) as
the functional of the order parameter $\phi=\phi({\bf r})=\langle
\hat\psi({\bf r}) \rangle$ (here $\hat\psi({\bf r})$ is the Bose
field operator)
\begin{equation}
E = \int d^3r\left[\frac{\hslash^2}{2m}\,|\nabla \phi|^2+V_{\rm
ext}({\bf r}) |\phi|^2 +\frac{2\pi\hslash^2a}{m} |\phi|^4 \right],
\label{FGP}
\end{equation}
called the Gross-Pitaevskii functional. In Eq.~(\ref{FGP}) $V_{\rm
ext}({\bf r})$ denotes an external trapping field with the
characteristic length $L$, typical for the spatial variation of this
field [for a harmonic trap, $L\simeq a_{ho} = \sqrt{\hslash /(m
\omega_{ho})}\,$], and $|a| \ll L$ stands for the scattering length
in a gas considered. From the definition it follows that the order
parameter satisfies the normalization condition $\int d^3r
|\phi|^2=N$, where the number of bosons $N_{0}$ in the Bose-Einstein
condensate is replaced by their total number $N$ due to a small
condensate depletion. The stationary solution corresponding to the
minimum of the functional (\ref{FGP}) obeys the relations
\begin{equation}
\delta(E-\mu N)/\delta\phi^{*}({\bf r})=\delta(E-\mu N)/
                                        \delta\phi({\bf r})=0,
\label{mincond}
\end{equation}
which give the GP equation
\begin{equation}
\mu\phi=-\frac{\hslash^2\nabla^{2}}{2m}\phi+V_{\rm ext}({\bf r})\phi
+\frac{4\pi\hslash^2a}{m}|\phi|^2\phi.
\label{GPE}
\end{equation}
The term $\mu N$ appears in Eq.~(\ref{mincond}) due to the
normalization condition, and the Lagrange multiplier $\mu$ is nothing
but the chemical potential. In this paper we address {\it the problem
of interpretation of different terms in the GP functional; namely,
the kinetic $E_{\rm kin}$ and pairwise interaction $E_{\rm int}$
energies are under investigation below}. Note that this is of actual
interest because analysis of the experiments with magnetically
trapped Bose gases is carried out usually in terms of $E_{\rm kin}$
and $E_{\rm int}$~\cite{RMP,Ket}. Moreover, the problem of that
interpretation is rather ambiguous due to just opposite points of
view on this subject found in the literature.  Indeed, in the
paper~\cite{Ket} the third term of the GP functional is treated as
the energy of the pairwise boson interaction. While according to
another point of view~\cite{Lieb1}, in the homogeneous case $V_{\rm
ext}=0$ this term makes contribution only to the kinetic energy if
the boson-boson interaction potential $V(r)$ is of the hard-sphere
form: $V(r)=+\infty$ at $r < a$, and $V(r)=0$ at $r \geqslant a$.


This problem is usually considered to be trivial~\cite{RMP}. Indeed,
in the mean-field interpretation~\cite{Note1}, the terms in
Eq.~(\ref{FGP}) are associated with the kinetic energy, the energy of
interaction with the external field, and the pairwise interaction
energy, respectively. However, in the particular case $V_{\rm
ext}=0$, this interpretation contradicts both the results of Lieb and
Yngvason~\cite{Lieb1} and ours~\cite{our}. To have an understanding
of the situation, let us calculate the kinetic and interaction
energies of a trapped Bose gas on a solid theoretical basis beyond
the mean-field interpretation of the terms of the GP functional. This
can be realized with the help of the well-known variational theorem
for the grand canonical potential at zero temperature
$\delta\Omega=\delta(E-\mu N)= \langle \delta(\hat{H} -
\mu\hat{N})\rangle$, where $\langle\cdots\rangle$ stands for the
statistical average over the grand canonical ensemble. To calculate
the pairwise interaction energy $E_{\rm int}= \langle \sum_{i\not=j}
V(|{\bf r}_{i} -{\bf r}_{j}|)/2\rangle$, it is useful to introduce
the coupling constant $\lambda$ [$V(r) \to \lambda V(r)$, and
$\lambda=1$ in final formulas]. Then the thermodynamic variational
theorem leads to $E_{\rm int}=\partial (E-\mu N) /
\partial \lambda$, which, in conjunction with Eqs.~(\ref{FGP}) and
(\ref{mincond}), yields~\cite{analit}
\begin{equation}
E_{\rm int} =\frac{1}{2}\Big\langle\sum_{i\ne j}V(|{\bf
r}_{i}-{\bf r}_{j}|) \Big\rangle
=\frac{2\pi\hslash^2}{m}\frac{\partial a}{\partial \lambda}\int
d^3r\,|\phi|^4, \label{Eint1}
\end{equation}
where $\phi$ is the stationary order parameter obeying the GP
equation (\ref{GPE}). Being the characteristic of the two-body
problem, the derivative $\partial a/\partial \lambda$ is obtained
from the other variational theorem proved in the papers~\cite{our}
for the short-range pairwise interaction potentials that go to zero
at $r \to \infty$ as $V(r)\to 1/r^{m}$ ($m>3$), or even faster. The
particular variant of this theorem allows for connecting the
infinitesimal change of the scattering length $\delta a$ with that of
the potential $\delta V(r)$:
\begin{equation}
\delta a=\frac{m}{4\pi\hslash^2}\int d^3r\,[\varphi^{(0)}(r)]^{2}
\delta V(r), \label{TCh}
\end{equation}
where $\varphi^{(0)}(r)$ is the s-wave solution of the two-body
Schr\"odinger equation in the center-of-mass system
\begin{equation}
-(\hslash^2/m)\nabla^2\varphi^{(0)}(r)+V(r)\varphi^{(0)}(r)=0.
\label{twobody}
\end{equation}
Equation (\ref{twobody}) corresponds to the scattering state with the
momentum $p=0$, and its solution $\varphi^{(0)}(r)$ is chosen to be
real due to the real boundary condition $\varphi^{(0)}(r)\to 1-a/r$
at $r \to \infty$. This boundary condition and Eq.~(\ref{twobody})
yield the following relation:
\begin{equation}
\frac{4\pi\hslash^2 a}{m}=\int d^3r\,V(r)\varphi^{(0)}(r).
\label{a}
\end{equation}
Equations~(\ref{TCh}) and (\ref{twobody}) lead to
\begin{equation}
\lambda\frac{\partial a}{\partial \lambda}=\frac{m}{4\pi\hslash^2}
\int d^3r\,\lambda V(r)[\varphi^{(0)}(r)]^2 = a-b, \label{dadl}
\end{equation}
where we put by definition
\begin{eqnarray}
b=\frac{1}{4\pi}\int
d^{3}r\,\bigl|\nabla\varphi^{(0)}(r)\bigr|^{2}. \label{b}
\end{eqnarray}
From Eq.~(\ref{b}) it follows that $b$ is a positive quantity and can
be considered as a new characteristic length, which is not expressed
in terms of $a$ and depends on a particular shape of the interaction
potential $V(r)$. For example, for the hard spheres [$V(r)=+\infty$
at $r < a$, and $V(r)=0$ at $r \geqslant a$] we have
$b=a$~\cite{our}. While for $V(r)$ close to zero, in the
weak-coupling regime~\cite{classif}, we obtain $b \ll
|a|$~\cite{our}. Thus, Eq.~(\ref{Eint1}), taken together with
Eq.~(\ref{dadl}), gives
\begin{equation}
E_{\rm int}=\frac{2\pi\hslash^2}{m}(a-b)\int d^3r\,|\phi|^4.
\label{Eint2}
\end{equation}
In the same manner, using the thermodynamic variational theorem
with respect to $\lambda V_{\rm ext}({\bf r})$, one obtains for
the energy of interaction with the external field
\begin{equation}
E_{\rm ext} =\Big\langle\sum_{i} V_{\rm ext}({\bf
r}_{i})\Big\rangle = \int d^3r\,V_{\rm ext}({\bf r})|\phi|^2.
\label{Eext}
\end{equation}
In turn, the kinetic energy can be calculated by using Eqs.~(\ref{FGP}),
(\ref{Eint2}) and (\ref{Eext}) with the result
\begin{equation}
E_{\rm kin}
\!=\!\Big\langle\sum_{i}\frac{p_{i}^{2}}{2m}\Big\rangle\!
=\!\int\!d^3r\! \left[\frac{\hslash^2}{2m}|\nabla \phi|^2
+\frac{2\pi\hslash^2 b}{m}|\phi|^4 \right]. \label{Ekin2}
\end{equation}
Thus, the thermodynamic variational theorem and mean-field treatment
[the latter leads to Eqs.~(\ref{Eint2})-(\ref{Ekin2}) but with $b=0$]
yield in general different results for the kinetic and pairwise
interaction energies of a trapped Bose gas. Here one should not be
confused by the fact that the GP equation is usually considered to be
a product of the mean-field approach. Though it has originally been
derived by means of this approach~\cite{GP}, there exists the
rigorous derivation of the GP functional~\cite{Lieb2} beyond the
mean-field approximation. Worth mentioning that this approximation
has recently been demonstrated~\cite{our} to face significant
problems in the strong-coupling regime~\cite{classif}. The point is
that the mean-field approach yields the correct result for the total
energy of a Bose gas of strongly interacting particles (strictly
speaking, after removing the ultraviolet divergence~\cite{our}).
However, it fails to give the appropriate picture of the short-range
boson correlations in this case, which leads to the ultraviolet
divergence and incorrect values for the kinetic and interaction
energies~\cite{our}. We stress that this only concerns the
strong-coupling case. There is no problem with the mean-field theory
in the weak-coupling regime, when $b/|a| \ll 1$, and, hence, one can
approximately put $b=0$ in Eqs.~(\ref{Eint2}) and (\ref{Ekin2}). Note
that in the uniform case $V_{\rm ext}=0$ the GP equation (\ref{GPE})
gives $\phi={\rm const}=\sqrt{n}$ (here $n=N/V$ is the density of
bosons), and Eqs.~(\ref{Eint2}) and (\ref{Ekin2}) are reduced to the
results
\[
 E_{\rm int}/N = 2\pi\hslash^2 n(a-b)/m,\quad
 E_{\rm kin}/N = 2\pi\hslash^2 n b/m,
\]
derived earlier by the present authors~\cite{our}.

Mathematically rigorous, the derivation~\cite{Lieb2} of the GP
functional does not give any information about nature of the third
term in Eq.~(\ref{FGP}). This nature becomes, of course, clear from
Eqs.~(\ref{Eint2})-(\ref{Ekin2}), but the method of obtaining them is
also rather formal. For a more deep insight, let us find out what
approximations for the reduced density matrices result in
Eqs.~(\ref{Eint2}) and (\ref{Eext}). Introducing the one-body
(1-matrix) $F_1({\bf r},{\bf r}')= \langle \hat\psi^{ \dagger} ({\bf
r}) \hat\psi({\bf r}')\rangle$ and two-body (2-matrix) $F_2({\bf
r}_1,{\bf r}_2;{\bf r}_1', {\bf r}_2')= \langle
\hat\psi^{\dagger}({\bf r}_1)\psi^{\dagger} ({\bf r}_2) \hat\psi({\bf
r}_2')\hat\psi({\bf r}_1') \rangle$ density matrices, one writes
\begin{eqnarray}
E_{\rm ext}&=& \int d^3r\,V_{\rm ext}({\bf r})F_1({\bf r},{\bf
r}),
\label{Eext3}\\
E_{\rm int}&=&\frac{1}{2} \int d^3r d^3r'\,V(|{\bf r}- {\bf r}'|)
F_2({\bf r},{\bf r}';{\bf r},{\bf r}'). \label{Eint3}
\end{eqnarray}
The 1-matrix can be expanded in the orthonormal set of its
eigenprojectors as
\begin{equation}
F_1({\bf r},{\bf r}')=N_{0}\phi_{0}^{*}({\bf r})\phi_{0}({\bf r}')+
\sum_{i\not=0} n_{i}\phi_{i}^{*}({\bf r})\phi_{i}({\bf r}').
\label{F1exact}
\end{equation}
Here $\phi_{0}({\bf r})=\phi({\bf r})/\sqrt{N_0}=
\langle\hat\psi({\bf r})\rangle/\sqrt{N_{0}}$ is the eigenfunction
corresponding to the macroscopic eigenvalue $N_{0}$ [$\int d^3r'
\, F_1({\bf r}',{\bf r})\phi_{0}({\bf r}') = N_{0} \phi_{0}({\bf
r})$, $\int d^3r\,|\phi_{0}({\bf r})|^{2}=1$], see, e.g.,
Ref.~\cite{RMP}.  By definition, the one-body matrix is normalized
as $\int d^3r \,F_1({\bf r},{\bf r})=N$; therefore, in the case of
a small condensate depletion $(N-N_{0})/N\ll 1$ one can put
\begin{eqnarray}
F_1({\bf r},{\bf r}')\simeq
\langle \hat\psi^{\dagger}({\bf r})\rangle\langle\hat\psi({\bf r}')
\rangle,
\nonumber
\end{eqnarray}
which, taken together with Eq.~(\ref{Eext3}), results immediately
in Eq.~(\ref{Eext}). Note that the calculation of the second term
of the kinetic energy~(\ref{Ekin2}) which involves the 1-matrix
requires knowledge of the other eigenvalues and eigenfunctions of
$F_1({\bf r},{\bf r}')$. However, this calculation is not really
needed because, when treating a trapped Bose gas on the basis of
the microscopical information concerning the one-body and two-body
matrices, we are able to restrict ourselves only to derivation of
Eqs.~(\ref{Eint2}) and (\ref{Eext}). As to Eq.~(\ref{Ekin2}), it
can be determined from the relation $E=E_{\rm kin}+E_{\rm
ext}+E_{\rm int}$ with the help of Eqs.~(\ref{FGP}),
(\ref{Eint2}), and (\ref{Eext}). Let us repeat one more that the
second term in Eq.~(\ref{Ekin2}) is exactly due to the second term
in Eq.~(\ref{F1exact}). For this reason, the approximation for the
momentum distribution $n_{\bf p}\simeq |\phi({\bf p})|^{2}$ (where
$\phi({\bf p})$ is the Fourier transform of the order parameter)
is able to produce only the first term in Eq.~(\ref{Ekin2}).

Explicit expressions for a number of eigenvalues and
eigenfunctions of the two-body matrix have been obtained for a
homogeneous system of bosons in the paper~\cite{Ch3} and were then
successfully employed in the papers~\cite{our}. In particular, the
anomalous two-operator (quasi-) average $\langle\hat\psi({\bf
r}_{1}) \hat\psi({\bf r}_{2}) \rangle/\sqrt{N_{0}(N_{0}-1)}$ was
shown to be the eigenfunction (pair wave function) of the
2-matrix, corresponding to the maximum macroscopic eigen\-value
$N_{0}(N_{0}-1)\simeq N_{0}^{2}$, which is nothing else but the
number of pairs of bosons in the Bose-Einstein condensate. This
result is also valid for a non-uniform (trapped) Bose system. In
the homogeneous case, this average, contrary to the situation for
the Fermi systems~\cite{Bog}, describes the {\it scattering} (not
bound) state of a pair of the condensed bosons. From this
interpretation it follows that at small separations the pair wave
function $\langle\hat\psi({\bf r}_{1}) \hat\psi({\bf
r}_{2})\rangle/\sqrt{N_{0}(N_{0}-1)}$ should be proportional to
the two-body wave function $\varphi^{(0)}(r)$ obeying the
Shr\"odinger equation (\ref{twobody}). On the other hand, the
Bogoliubov principle of the correlation weakening~\cite{Bog}
yields $\langle \hat\psi({\bf r}_{1})\hat\psi({\bf r}_{2})\rangle
\to \langle \hat\psi({\bf r}_{1})\rangle\langle\hat\psi({\bf
r}_{2})\rangle$ when $|{\bf r}_{1}-{\bf r}_{2}|\to \infty$. From
the physical point of view, the latter limit implies $|{\bf
r}_{1}-{\bf r}_{2}|\gg l_{\rm coh}$, where the length of
coherence~\cite{RMP} $l_{\rm coh}=1/\sqrt{8\pi n a}$ (here $n\simeq
|\phi({\bf r}_{1})|^{2}\simeq |\phi({\bf r}_{2})|^{2}$ is the
local density) is assumed to obey the inequalities $|a|\lesssim
l_{\rm coh}\ll L$. Therefore, we arrive at the following
approximation:
\begin{equation}
\langle\hat\psi({\bf r}_1)\hat\psi({\bf r}_2)\rangle\simeq\phi({\bf
r}_1) \phi({\bf r}_2) \varphi^{(0)}(|{\bf r}_1-{\bf r}_2|).
\label{psipsia}
\end{equation}
Remind that $\varphi^{(0)}(r)$ is the solution of
Eq.~(\ref{twobody}) with the asymptotics $\varphi^{(0)}(r)\to
1-a/r$ for $r \to \infty$. We stress that the approximation
(\ref{psipsia}) works well at $|{\bf r}_{1}-{\bf r}_{2}|\lesssim
l_{coh}$. By analogy with the 1-matrix, the small condensate
depletion leads to the representation
\begin{eqnarray}
F_2({\bf r}_1,{\bf r}_2;{\bf r}_1',{\bf r}_2')\simeq
\langle \hat\psi^{\dagger}({\bf r}_1)\psi^{\dagger}({\bf r}_2)
\rangle \langle\hat\psi({\bf r}_2')\hat\psi({\bf r}_1')\rangle,
\label{2red}
\end{eqnarray}
which, with the help of Eqs.~(\ref{Eint3}) and (\ref{psipsia}),
yields
\begin{eqnarray}
E_{\rm int}&=&\frac{1}{2} \int d^3r d^3r'\,V(|{\bf r}- {\bf r}'|)
[\varphi^{(0)}(|{\bf r}-{\bf r}'|)]^{2}
\nonumber \\
&&\times|\phi({\bf r})|^{2}|\phi({\bf r}')|^{2}.
\label{Eint4}
\end{eqnarray}
Finally, using the inequality $L\gg a$ valid in the
GP approximation  and the theorem (\ref{dadl}) at
$\lambda=1$, we arrive at Eq.~(\ref{Eint2}).

\begin{figure*}[t]
\centerline{\includegraphics[width=12.cm,clip=true]{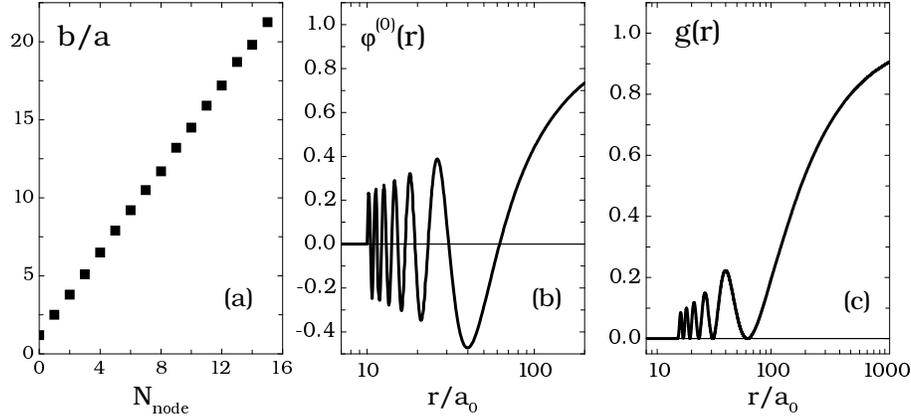}}
\caption{(a) The ratio $b/a$ versus the number $N_{\rm node}$ of
the nodes of $\varphi^{(0)}(r)$ [$\varphi^{(0)}(r)$ is the s-wave
solution of the Schr\"odinger equation~(\ref{twobody}) with $V(r)$
given by Eq.~(\ref{pot})] for the given values of the scattering
length $a=52 a_0$ and the model parameter $r_c=88.1 a_0$.  This
ratio is equal to $E_{\rm kin}/E$ for a homogeneous dilute Bose
gas of sodium atoms.  (b) The wave function $\varphi^{(0)}(r)$ for
the model potential~(\ref{pot}) with $N_{\rm node}=12$. (c) The
pair distribution function $g(|{\bf r}_{1}-{\bf r}_{2}|)=F_2({\bf
r}_1,{\bf r}_2; {\bf r}_1,{\bf r}_2)/[|\phi({\bf
r}_1)|^{2}|\phi({\bf r}_2)|^{2}]\simeq [\varphi^{(0)}(|{\bf
r}_{1}-{\bf r}_{2}|)]^{2}$ [see Eqs.~(\ref{psipsia}) and
(\ref{2red})] in a trapped sodium atoms, interacting via the
potential~(\ref{pot}) with $N_{\rm node}=5$.  Intensive
oscillations occur when $r \lesssim a=52 a_0$.} \label{fig1}
\end{figure*}

In order to estimate the ratio $b/a$ typical for the alkali atoms,
we make use of the model potential
\begin{equation}
V(r)=\left\{\begin{array}{ll}
+\infty,                      &r \leqslant r_0,\\
-\hslash^{2}r_{c}^{4}/(mr^6), &r > r_0.
\end{array}\right.
\label{pot}
\end{equation}
It is usually considered~(see, e.g., \cite{Gora}) to be relevant
in the situation of interest. This model interaction
leads~\cite{Mott} to the scattering length
\begin{equation}
a/r_c=\Gamma(3/4)J_{-1/4}(x_0)/\bigl[2\Gamma(5/4)
J_{1/4}(x_0)\bigr].
\label{aAL}
\end{equation}
Here $x_0=r_c^2/(2r_0^2)$; $J_{\nu}(x)$ and $\Gamma(z)$ denote the
Bessel function and the Euler gamma-function, respectively.
Further, calculating $\partial a/\partial \lambda$ with
Eq.~(\ref{aAL}) [$r_{c}\to\lambda^{1/4} r_{c}$ in Eq.~(\ref{pot})] and, then, using
Eqs.~(\ref{dadl}) and (\ref{aAL}) at $\lambda=1$, we arrive at
\begin{equation}
b/a=3/4+1/\bigl[\pi \sqrt{2} J_{1/4}(x_0)J_{-1/4}(x_0)\bigr].
\label{bAL}
\end{equation}
For the potential (\ref{pot}), the parameter $x_{0}$ is connected
with the number of the bound states for the two-body problem: larger
the number, larger $x_0$.  It is well-known that this number is equal
to that of the nodes of the wave function $\varphi^{(0)}(r)$ obeying
Eq.~(\ref{twobody}). The pairwise interaction of alkali atoms is
characterized by large amount of possible bound pair states, which,
however, kinetically inaccessible in the trapped systems~\cite{Ties}.
So, we are interested in the regime when $x_0 \gg 1$.  According to
Eq.~(\ref{bAL}), $b/|a| \gg 1$ at $x_0 \gg 1$. The same situation
($b/|a| \gg 1$) takes place also at any $x_{0}$ provided the
denominator in Eq.~(\ref{aAL}) is close to zero (as in the case of
the Feshbach resonance). The large value of $b/|a|$ means that the
second term in Eq.~(\ref{Ekin2}) is positive and much larger than the
absolute value of the third term in the GP functional, while the
pairwise interaction energy (\ref{Eint2}) is negative and close, in
absolute value, to the second term of Eq.~(\ref{Ekin2}). In other
words, for a homogeneous dilute Bose gas of alkali atoms [when the
first term in Eq.~(\ref{Ekin2}) equals to zero] we can expect $E_{\rm
kin} \simeq |E_{\rm int}| \gg |E|$. In order to obtain numerical
estimations for sodium atoms, we utilize the data of Ref.~\cite{Ties}
$r_c=88.1 a_0$ and $a=52 a_0$~($a_0$ is the Bohr radius).  Note that,
at any $r_c$ and $a$, solutions of Eq.~(\ref{aAL}) for the core
radius $r_0$ form an infinite sequence, approaching zero; each of the
solutions for $r_{0}$ corresponds to a certain number $N_{\rm node}$
of the nodes of $\varphi^{(0)}(r)$. Numerical results for $b/a$
versus $N_{\rm node}$ are shown in Fig. \ref{fig1}(a).
A typical solution of the Schr\"odinger equation (\ref{twobody}) with
the potential (\ref{pot}) corresponding to $N_{\rm node}=12$ [and,
hence, $r_0=10.03 a_0$] is given in Fig. \ref{fig1}(b). Now, the
Na-Na interaction is expected~\cite{Ties} to provide more than 15
possible bound states, and, thus, we have the estimation $b/a\gtrsim
20$ [see Fig.~\ref{fig1}(a)]. We remark that the visible linear
dependence of $b/a$ on the number of the nodes is a specific feature
of the attractive interatomic potential proportional to $1/r^6$ but
not a general relation between $b$ and $a$. We also stress that the
pair distribution function of sodium atoms given in
Fig.~\ref{fig1}(c) is equal to zero at $r=0$, contrary to the
expectations~\cite{Ket} inspired by the mean-field treatment.

In conclusion, the kinetic and interaction energies
(\ref{Eint2})-(\ref{Ekin2}) of a 3D dilute ground-state Bose
gas confined in a trap have been derived beyond the mean-field
interpretation. The pairwise interaction energy $E_{\rm int}$ is
found to be controlled by the two characteristic lengths: in addition
to the well-known scattering length $a$ [Eq.~(\ref{a})], the
expression (\ref{Eint2}) involves the positive parameter $b$
[Eq.~(\ref{b})]. Whereas the kinetic energy (\ref{Ekin2}) depends on
the pairwise interaction only through the characteristic length $b$.
The derived estimations suggest that in the experimentally
interesting case of a dilute Bose gas of alkali atoms the part of the
kinetic energy coming from the pairwise interaction is much larger
than the absolute value of the third term in the GP
functional~(\ref{FGP}).

Note that experimental observation of $E_{\rm kin}$ and $E_{\rm
int}$ is quite possible provided the diagonal element of the
2-matrix $F_2({\bf r},{\bf r}';{\bf r}, {\bf r}')$~(the familiar pair
distribution function) and the momentum occupation number 
$n_{\bf p}=\langle a^{\dagger}_{\bf p}a_{\bf p}\rangle$ are
experimentally determined. Indeed, the pairwise interaction energy
is then calculated by means of Eq.~(\ref{Eint3}) with the pairwise potential
$V(r)$ that can be found numerically (see, e.g., Ref.~\cite{Ties}). The
kinetic energy per particle can be represented as
\begin{eqnarray}
E_{\rm kin}/N=\int \frac{d^{3}p}{(2\pi)^{3}}\frac{\hslash^{2}p^{2}}{2m}
\frac{n_{p}}{n}.
\label{ekinexp}
\end{eqnarray}
The quantities $F_2$ and $n_{\bf p}$ are the well-known subjects of experimental
investigations in the condensed matter physics, and there is
technique of measuring these quantities (see, e.g., discussion in Ref.~\cite{zamb}).
However, in the 
situation of interest, the trouble point is that, in order to calculate $E_{\rm
kin}$, one needs to know $n_{\bf p}$ beyond the phonon regime for $p \gtrsim
\hslash/a$ because the contribution of the phonon region is small in the integral
in Eq.~(\ref{ekinexp}). Moreover, measurements
with sufficiently small step in the momentum direction are of importance
here to integrate properly in Eq.~(\ref{ekinexp}). The same is related to 
Eq.~(\ref{Eint3}), where the short-range behaviour $|{\bf r}-{\bf r}'|\lesssim a$
of the pair distribution function
$F_2({\bf r},{\bf r}';{\bf r}, {\bf r}')$ is of importance. For this reason, accurate
data are needed for the static
structure factor at $p \gtrsim \hslash/a$. Moreover, one again needs to make fine
step-by-step measurements. To the best of our knowledge, so far there are
only several experimental points for the static structure
factor~\cite{str}. The situation concerning $n_{\bf p}$ is now even worse. Thus,
additional experiments are needed to make final conclusions about the
failure of the mean-field approach.

It is also worth noting that, in experiments on the expansion of
a condensate, one observes a kinetic energy of the freely expanding
condensate, which is equal to a {\it sum} of the pairwise potential
(\ref{Eint2}) and kinetic (\ref{Ekin2}) energies of the trapped one,
in effect proportional to the scattering length.

This work was supported in part by the RFBR grant 00-02-17181.



\begin{references}
\bibitem{analit} We assume that, mathematically, the point $\lambda=1$ is
a regular point for both the ground state energy and its
asymptotics -- the GP functional (\ref{FGP}). From
physical point of view, it means that there is no phase transition
at $\lambda=1$, associated with ``switching on" the coupling
constant $\lambda$, or another singularity like the Feshbach
resonance at that point.
\bibitem{classif} We employ the following classification of the short-range
pairwise potentials. The potential is of the weak-coupling type if it is
integrable and proportional to a small parameter~(the coupling constant). In
this case the Born approximation works well when applied to the ordinary
two-body problem. Otherwise the potential is called singular. Finally, we
deals with the strongly singular potential (the strong-coupling regime) if
the individual terms in the Born series for the scattering amplitude do not
exist due to strong repulsion at short particle separations. For example, the
Lennard-Jones and hard-sphere potentials belong to that type.
\end{references}
\end{document}